\documentclass[a4paper,12pt]{article}
\usepackage{xcolor}
\usepackage{amsfonts}
\usepackage{makeidx}
\usepackage{amsmath,amssymb,bbm} 
\usepackage{graphicx} 
\usepackage{cite} 
\usepackage[multiple]{footmisc}
\usepackage[utf8]{inputenc}

\usepackage[colorlinks=true,
linkcolor=red,
urlcolor=blue,
citecolor=blue]{hyperref}

\newcommand\eqs[1] {\begin{align}#1\end{align}}

\newcommand\eqsc[1] {\begin{gather}#1\end{gather}}
\newcommand\eqscn[1] {\begin{gather*}#1\end{gather*}}

\newcommand{\bea}{\begin{eqnarray}}
\newcommand{\eea}{\end{eqnarray}}

\DeclareMathOperator{\Tr}{Tr}

\newcommand\A {{\cal A}}

\renewcommand\L {{\cal L}}

\newcommand\N {{\cal N}}

\newcommand\V {{\cal V}}

\newcommand\ie {\textit{i.e.~}}
\newcommand\eg {\textit{e.g.}}

\newcommand{\viz}{viz.~}

\newcommand{\nl}{\newline}

\title{\bf Supersymmetric Localization on dS: Sum Over Topologies}
\author{\bf Rudranil Basu$^{a, b}$\thanks{rbasu@g.harvard.edu}, Augniva Ray$^c $\thanks{augniva.ray@saha.ac.in}
	\\ \medskip \\
	{}$a)$ \em  Center for the Fundamental Laws of Nature, Harvard University, \\ \em Cambridge, MA 02138, USA \\
	{}$b)$ \em BITS-Pilani, KK Birla Goa Campus, NH 17B, Bypass Road, \\ \em Zuarinagar, Goa, India 403726 \\
	{}$c)$ \em Theory Division, Saha Institute of Nuclear Physics, HBNI,\\ \em 1/AF Bidhan Nagar, Kolkata, India 700064}

\date{}

\begin{document}
	
	\maketitle

	\begin{abstract}
		We find the exact quantum gravity partition function on the static patch of 3d de Sitter spacetime. We have worked in the Chern Simons formulation of 3d Gravity. To obtain a non-perturbative result, we supersymmetrized the Chern Simons action and used the technique of supersymmetric localization. We have obtained an exact non-perturbative result for the spin-2 gravity case. We comment on the divergences present in the theory. We also comment on higher spin gravity theories and analyse the nature of divergences present in such theories.
	\end{abstract}

\bigskip
	
\section{Introduction}
Quantum theory of gravity in 3 space-time dimensions does not cease to surprise us, owing to the richness of physical and mathematical structures that are being continually revealed for more than 3 decades starting from \cite{WITTEN198846}. It is interesting that, gravity in 3 dimensions is devoid of local degrees of freedom. One of the main causes of non-triviality in 3D gravity is the BTZ black hole solution \cite{Banados:1992wn} for negative cosmological constant. The most interesting sector of solutions for the case of negative cosmological constant is asymptotically AdS. A huge body of work has stemmed from the seminal work of Brown and Henneaux \cite{Brown:1986nw}, which showed that the asymptotic symmetries of asymptotically AdS space-time form two copies of Virasoro algebra; thereby hinting to a plausible conformal field theory (CFT) at the two dimensional asymptotic boundary. As an example of low dimensional holography, this generated a great deal of physical and mathematical curiosities; motivated just from the question of calculating partition function for quantum gravity and arriving at black hole entropy from it. Interested readers may refer Refs. \cite{Maloney:2007ud,Belin:2018oza},  in recent times.

Analogous progress in the case for zero cosmological constant is being pursued recently, specially in the works of Refs. \cite{Barnich:2015mui,Bagchi:2019unf} . In this sector, one attempts at quantum gravity for asymptotically flat space-time, now equipped with the BMS${}_3$ algebra. \cite{Prohazka:2017lqb} contains a relatively extensive discussion of quantum gravity in 3 dimensions from the perspective of asymptotic symmetries for asymptotically non-AdS space-time, even including higher spin degrees of freedom.

Whereas these aspects of quantum gravity are under focus of intensive studies in recent times, one might be curious for the case of positive cosmological constant. Vacuum solution to the corresponding Einstein equation is the dS${}_3$ space-time. However unlike Minkowski space-time, here exists a horizon at thermal equilibrium. As argued in \cite{Castro:2011xb}, correlation function of any quantum degree of freedom with respect to a time-like observer is a thermal correlator. The corresponding vacuum state, as discussed in \cite{PhysRevD.28.2960} and named as the Hartle Hawking state, is the Euclidean partition function. 

The choice of Hartle Hawking state as a candidate for vacuum state circumvents an otherwise conceptually difficult problem in the following manner. Standard wisdom says that isometries of a maximally symmetric space-time like de Sitter should fix the vacuum state. But if one wishes to incorporate effects from quantum gravity, one has to incorporate all possible quantum fluctuations on the de Sitter background, from a perturbative viewpoint. Hartle Hawking is however defined as the Euclidean path integral considering all possible geometries with some fixed boundary data. 

Now in de Sitter space, a time-like observer is in causal contact with what is known as the static patch, defined in Euclidean time as:
\begin{eqnarray}\label{met1}
ds^2 =dr^2 + \cos^2 r d \tau^2 + \sin^2 r  d\phi^2.
\end{eqnarray}
Euclideanizing is done by setting $t = -i \tau$ and it makes the static patch geometry identical to that of $S^3$ with $\tau \in [0, 2\pi], \phi \in [0, 2\pi], r \in [0, \pi/2]$. 

It would therefore be natural to consider fluctuations over round $S^3$ background geometry to construct the Hartle Hawking state. However, as nicely pointed out in \cite{Castro:2011xb}, there is an infinite class of topologically distinct manifolds which allow smooth local geometry as Eq. \eqref{met1}. These are of the form $S^3/\Gamma$, where $\Gamma$ is a discrete subgroup of the isometry group of $S^3$. In terms of the coordinates in Eq. \eqref{met1}, these quotient spaces with smooth local dS geometry are understood by the following identifications:
\begin{eqnarray} \label{ilens}
(\tau, \phi) \sim (\tau, \phi) + 2 \pi\,(\frac{m}{p}, \frac{m\,q}{p} + n) \quad \mbox{ for } \quad m,n \in \mathbb{Z}.
\end{eqnarray}
Here $q,p$ are coprime positive integers with $p$ always being the greater of the pair. That this identification indeed results into the topological quotient space $S^3/ \mathbb{Z}_p$ can be easily understood by first defining 
\begin{eqnarray}
S^3 = \{ (z_1, z_2) \in \mathbb{C}^2 \Big{|} |z_1|^2 + |z_2|^2 =1\}. 
\end{eqnarray}
Then the $\mathbb{Z}_p$ action on it is:
\begin{eqnarray}
\left(z_1, z_2\right) \rightarrow \left(e^{\frac{2 \pi i}{p}} z_1, e^{\frac{2 \pi i q}{p}} z_2\right). 
\end{eqnarray}
Finally defining $ (z_1, z_2) = \left(\cos r \,e^{i \tau}, \sin r \, e^{i \phi} \right)$ makes the identification Eq. \eqref{ilens} clear. The resultant manifold is named as a Lens space $L(p,q)$, now equipped with the smooth geometry given by Eq. \eqref{met1}. All of these manifolds are therefore valid classical smooth saddles of Einstein equation.

Since $S^3$ as well as all the quotients $L(p,q)$ are closed, Hartle Hawking state, considering all quantum gravity effects, would simply be given by:
\begin{eqnarray}
Z = \sum_{L(p,q)} \int [Dg] e ^{-S_E[g]} \label{partitionfunction}
\end{eqnarray}
where $S_E$ is the Euclidean action for the theory of gravity. Interestingly as reported in \cite{Castro:2011xb}, the functional integral, when summed over all Lens space saddles, diverges as a harmonic series in the integer $p$: $ \sum_{p=1}^\infty \frac{1}{p} = \zeta(1)$, which cannot be regularized. The computation for a single Lens space (before summing up) was performed in a perturbative one-loop calculation in metric variables and cross-checked with results from a non-perturbative computation in first order formulation of gravity (Chern Simons (CS) theory) \cite{jeffrey1992}. One might not wish to compare this divergence to the well known divergences regularly encountered in QFTs. Firstly because, this model, unlike standard QFTs is devoid of local degrees of freedom. The second and more subtle issue is that, the full quantum theory of gravity should be background independent and hence should not have any canonical definition of energy scale. The absence of local physics, makes it difficult to understand the origin, and hence controlling the divergence becomes a very hard problem.

However, the divergence seems to be tamed, when including further degrees of freedom, like topological massive modes \cite{Castro:2011ke} making the Hartle Hawking vacuum state normalizable. This was later established \cite{Basu:2011vs} using a twisted first order theory of gravity (again CS formulation) and a dimensionless parameter, which can be tuned to get rid of the divergent piece. Interestingly, using results from $SU(N)$ topological invariants \cite{doi:10.1142/S0218216504003342} in 3-manifolds one can repeat the calculations for higher spin cases. For this, one introduces a consistently truncated tower of higher spins over gravitational degrees of freedom, to see that the sum over all Lens spaces become finite, for spins $\geq 4$ \cite{Basu:2015exa}. Even with these attempts, it is still not clear, which deformations or coupling of newer excitations on top of gravity can make the Hartle-Hawking vacuum state normalizable. It therefore calls for further attempts to make a classification scheme for such well behaving excitations, in a fashion analogous to classifying renormalizable quantum field theories. However let us emphasize that the removal of divergences mentioned above by introduction of newer degrees of freedom is not equivalent to renormalization via addition of counter-terms in QFT since the divergence is not caused by local physics.

One further motivation towards a definition of Hartle Hawking state in 3D quantum de Sitter comes from an analogous question in AdS$_3$.
Euclidean AdS has a topology of solid torus. The two dimensional toric boundary serves as the asymptotics. Using the fact that asymptotic symmetry in AdS${}_3$ is given by 2D conformal algebra, one may come up with speculations \cite{Castro:2011zq} regarding a candidate 2D CFT at the boundary. An exact non-perturbative calculation for the bulk partition function (corresponding to fixed boundary modular parameter) can lead one a long way towards a definite answer regarding the dual field theory. A series of recent remarkable results in AdS, Refs. \cite{Iizuka_2015,Honda_2017} have taken the approach of supersymmetrizing the gravity theory (CS formulation) and exploiting the elegant methods of supersymmetric localization \cite{Pestun:2007rz}.  Although the original theory, modelled as a purely bosonic theory of gauge fields, the localization procedure brings in fermionic degrees of freedom in the dynamics. However, it remains guaranteed, as we'll later review	in the present article that the computation of the partition function for the localized theory is same as the one, if one could evaluate the one for the original purely bosonic one. In the AdS case it is believed that the non-perturbative result after localization would constrain completely the CFT dual to the original bosonic gravity theory. For further progress in localization in low-dimensional AdS space times, the interested reader is referred to Refs. \cite{David:2016onq, Assel:2016pgi}. These references focus on the program of localization on non-compact manifolds.

In our present perspective we don't aim at the holographic point of view. But rather take cue from the above analysis as far as exact partition function is concerned. We want to investigate if the divergence in Hartle-Hawking state, previously  found in purely bosonic theory, while summing over all saddles can be tamed or modified by the introduction of supersymmetry. To this end  we use the first order CS formulation here, and supersymmetrize it to write down the exact partition function. 

To put this point properly in context of our present work, let us digress a bit on the meaning of partition function. In traditional quantum field theory on a fixed background (Minkowski space, for example) describing fundamental interactions, partition function is an extremely efficient tool for evaluating correlation functions. As long as the quantum theory of a classical Euclidean action $S_E[\Phi]$ is renormalizable, one is generally interested in the functional integral:
\bea{}
Z[J] = \int D\Phi \, e^{-S_E[\Phi] + \int J \cdot \Phi}
\eea
in presence of a probe background field $J$. Correlations of local operators
\bea{} \label{corr}
\langle O_1 (x_1) \cdots O_n (x_n) \rangle = \frac{\int D\Phi \, O_1 (x_1) \cdots O_n (x_n) \, e^{-S_E[\Phi] }}{Z[0]}
\eea
generally are found as appropriate derivatives of $Z[J]$ with respect to $J$ at the point $J=0$, while the presence of `normalizing' factor $Z[0]$ in the denominator of \eqref{corr} is also standard.

In contrast, the goal of the present series of works \cite{Castro:2011xb}, \cite{Castro:2011ke}, \cite{Basu:2011vs}, \cite{Basu:2015exa} is to investigate the Hartle-Hawking vacuum via evaluating the partition functions of Chern Simons theory considering all saddles relevant to gravity on locally de Sitter background and then sum over all geometries. These saddles being Lens spaces, each such partition function is a topological invariant \cite{doi:10.1142/S0218216504003342} and for each Lens space $L(p,q)$, the CS partition function is a functions of $p,q$ and the CS level $k$ (possible shifted by quantum correction). We will see in our analysis, how this quantity for each Lens space, has a spin-dependent multiplicative power law dependence on the volume of that particular Lens space. We will notice in this article that fermions brought in by supersymmetric localization basically does the job of altering the volume prefactor's power, keeping the rest of the functional dependence of $k,p, q$ unchanged with respect to the bosonic case. This alteration introduced by fermions, makes the previously encountered divergence worse.

However, it is fair to assume that, had we been interested in a fixed background question of calculating correlators \footnote{Since pure Einstein gravity in 3 dimensions is devoid of local dynamics, it is hard to define physical, non-trivial correlation functions, particularly in the bulk.} via the definition \eqref{corr} on a fixed $L(p,q)$, the prefactor would have got cancelled due to normalization and the results would remain same as in the purely bosonic theory.

Furthermore, an investigation for whether inclusion of higher bosonic spins and the corresponding supersymmetrization would change the behaviour of the proposed partition function is also due. We here realize a better insight into the interplay between the spin content in the theory and the divergence structure. In previous  analysis in Ref. \cite{Basu:2015exa}, it was encountered that bosonic higher spin contributions make product of volume prefactors suppress the divergent contributions. We will elaborate here the quite opposite feature here.

As choice of newer degrees of freedom, higher spins are obvious, as these in 3 dimensions are much more tractable than in the case of higher dimensions, because of an allowed consistent truncation of the higher spin tower at any finite spin $ > 2$. Effect of finite number of higher spin fluctuations coupled with the background spin-2 fluctuations have been as found in numerous AdS and flat-space calculations. Analysis in the presence of higher spin fields in AdS spacetimes has been worked out in the seminal works by Gopakumar et al in Refs. \cite{Gaberdiel:2014cha,Gaberdiel:2015mra,Gaberdiel:2015wpo,Gaberdiel:2017dbk,Gaberdiel:2010ar}. In flat spacetimes, similar such work has been carried out in Refs. \cite{Ammon:2017vwt,Gary:2014ppa,Afshar:2013vka}.

The paper is organized as follows. In section \ref{thetheory}, we introduce the CS formulation of 3d bosonic gravity. In subsection \ref{supersymmetrization}, we obtain the supersymmetric extension of bosonic CS theory. In subsection \ref{Loc}, we discuss the technique of supersymmetric localization of our theory. In subsection \ref{partfunc}, we explicitly evaluate the partition function, obtained as a matrix model, for our case of spin-2 gravity. We also explicitly identify the divergent pieces in the partition function. In the following section \ref{hahah} , we evaluate the same for higher spin cases and comment on the divergences observed.  In section \ref{conclusions} , we comment on some future directions that may be explored.  Section \ref{appendix} carries a note on our definitions and conventions.

\section{Chern Simons formulation for 3d gravity and its supersymmetrization} \label{thetheory}

3D gravity is long known to be equivalent to a pure CS theory \cite{WITTEN198846}. Let us first briefly take a detour through this equivalence, particularly for the case of positive cosmological constant in Euclidean setting. One can start off with a CS functional on a 3-manifold $M$ out of a $\mathfrak{su}(2) \oplus \mathfrak{su}(2)$ valued 1-form (gauge field). Also the Lie algebra is equipped with an Ad invariant symmetric bilinear quadratic form $\mathrm{Tr} \equiv \langle \cdot, \cdot \rangle$ valued to be $\mathrm{diag}(k,k,k)$ and $\mathrm{diag}(-k,-k,-k)$ respectively on the first and the second $\mathfrak{su}(2)$. The CS functional then can be written as difference of two $\mathfrak{su}(2)$ CS functional, $\Tr$ now evaluating $\mathrm{diag}(1,1,1)$:
\begin{eqnarray}
\label{csaction}
S[\mathcal{A}^+ ,\mathcal{A}^-] &=& \frac{k}{4 \pi} \Tr\int _M ( \mathcal{A}^+ \wedge d\mathcal{A}^+ + \frac{2}{3} \mathcal{A}^+ \wedge\mathcal{A}^+ \wedge \mathcal{A}^+)\nonumber \\&-&\frac{k}{4 \pi} \Tr \int _M ( \mathcal{A}^- \wedge d\mathcal{A}^- + \frac{2}{3}\mathcal{A}^- \wedge\mathcal{A}^- \wedge \mathcal{A}^-)
\end{eqnarray}
with,
\begin{eqnarray}
\mathcal{A}^{\pm} = \omega \pm e~, \qquad k=\frac{1}{4G}  \label{gaugeflddef}
\end{eqnarray}  keeping the positive cosmological constant $\Lambda=1$, $G$ is the Newton's constant in 3 dimensions and $e$ and $\omega$ are the $\mathfrak{su}(2)$ triad and connection respectively. It is easy to see that Eq. \eqref{csaction} is actually the action for first order gravity:
\begin{eqnarray} \label{palatini}
\frac{k}{2\pi}\int_M  \left( e^I \wedge \left( 2 d \omega _I + \epsilon_{IJK} \omega ^J \omega ^K \right) + \frac{1}{3} \epsilon _{IJK} e ^I \wedge e^J \wedge e^K \right)
\end{eqnarray}
If $M$, is closed (for example the manifolds we will be dealing with in this article, ie the static patch of Euclidean dS${}_3 \sim S^3$ or $S^3 / \Gamma$), ie $\partial M = \emptyset $, the variational principle holds for the action Eq. \eqref{csaction} without any concern for boundary terms. 
Equations of motion are flatness conditions of the CS connections, ie $F^{\pm} = d \mathcal{A}^{\pm}$ which translate into
\begin{subequations} \label{eomc}
	\begin{eqnarray}
	&&\mathrm{torsionless ~condition:} ~~~~ d e^I + \epsilon ^{IJK} e_J \wedge \omega _K =0 \label{eoma}~~\mbox{and} \\
	&&\mathrm{curvature ~equation:} ~~~~ 2 d \omega ^I + \epsilon ^{IJK} \omega_J \wedge \omega _K = - \epsilon ^{IJK} e_J \wedge e _K \label{eomb}
	\end{eqnarray}
\end{subequations}
for gravity variables.
Interestingly, the following action 
\begin{eqnarray} \label{mod}
\tilde{S}[\mathcal{A}^+ ,\mathcal{A}^-] &=& \frac{k_+}{4 \pi} \Tr\int _M ( \mathcal{A}^+ \wedge d\mathcal{A}^+ + \frac{2}{3} \mathcal{A}^+ \wedge \mathcal{A}^+ \wedge \mathcal{A}^+)\nonumber \\&+&\frac{k_-}{4 \pi} \Tr \int _M ( \mathcal{A}^- \wedge d\mathcal{A}^- + \frac{2}{3} \mathcal{A}^- \wedge \mathcal{A}^- \wedge \mathcal{A}^-)
\end{eqnarray}
with independent levels $k_{\pm}$ also gives the same equations of motion Eq. \eqref{eomc} for gravity variables. For sake of convenience we introduce a parameter $\gamma$ such that, $ k_{\pm} = \frac{a(1/\gamma \pm 1)}{4G}$ and Eq. \eqref{mod} gives back Eq. \eqref{csaction} at the limit $\gamma \rightarrow \infty$ \cite{Basu:2011vs}. The equations of motion are independent of $\gamma$. This applies to the space solutions as well. On the other hand, other aspects of the dynamics of the theory, ie. canonical structures are parametrized by $\gamma$. For example, the pre-symplectic structure on the space of solutions Eq. \eqref{eomc}:
\begin{eqnarray}
\Omega(\delta_1,\delta_2)&=&\dfrac{k_+}{2\pi} \Tr \int_{\Sigma}\delta_1 \mathcal{A}^{+} \wedge \delta_2 \mathcal{A}^{+} - \dfrac{k_-}{2\pi} \Tr\int_{\Sigma}\delta_1 \mathcal{A}^{-} \wedge \delta_2 \mathcal{A}^{-} \nonumber \\&=& \dfrac{2a}{8 \pi G} \left(  \int_{\Sigma} (\delta_1 \omega^I \wedge \delta_2 \omega^I + \delta_1 e^I \wedge \delta_2 e^I) + \dfrac{2}{\gamma} \int_{\Sigma} \delta_{[1} \omega^I \wedge \delta_{2]} e^I  \right) 
\end{eqnarray}

Recall, in our definition, $A_{[a}B_{b]} \equiv \frac{1}{2}(A_a B_b - A_b B_a  )$

\subsection{Supersymmetrization} \label{supersymmetrization}
To evaluate the partition function given by Eq. \eqref{partitionfunction} exactly, we would use the recently developed supersymmetric Localization techniques of Pestun et al \cite{Pestun:2007rz}, adapted to our purpose. Towards this, we start by supersymmetrizing a CS gauge field $\mathcal{A}$ valued in some semi-simple Lie algebra. Later we will specialize to mainly $\mathfrak{su}(2)$, the case of relevance to 3D gravity. We construct the 3d $\N=2$ vector multiplet, defined, as always as $\V = (\A,\sigma, D,\lambda, \bar{\lambda})$.

The supersymmetric CS Theory action is written as
\eqsc{S_{SCS} [\V]=S_{CS}[\A] + \int d^3 x \sqrt{|g|} \ \Tr (-\lambda \bar{\lambda}+2 D \sigma)\label{7}} 
Note that in the 3d $\N=2$ vector multiplet, the additional fields $(\sigma, D,\lambda, \bar{\lambda})$ are not dynamical and give no kinetic terms contributions to the action. 
\section{Localization of the 3d Supersymmetric Chern Simons Theory on Lens Spaces}
With the connection between 3d Euclidean gravity and the Supersymmetric CS Theory made explicit in Eqs. \eqref{csaction} and \eqref{palatini}, we will now evaluate the partition function of the 3d Supersymmetric CS Theory via supersymmetric Localization techniques. 
Since, we are interested in evaluating gravity partition function on Lens Spaces, we would we would try localizing the CS Theory on Lens Spaces $L(p,q)$. 
\subsection{Principle of Localization} \label{Loc}
Suppose we have a theory on a compact manifold $\mathcal{M}$, defined by an action S$[\Phi]$\footnote{$\left\lbrace \Phi\right\rbrace $ stands for whatever the fields of the theory are.}, which has a Grassmann-odd symmetry\footnote{$\delta$ is assumed non anomalous. This is a crucial and non-trivial point.} $\delta$. Let us further assume that there exists an operator $V$ which is invariant under the transformation $\delta^2$, \ie $\delta^2 V=0$. Once we have established the existence of such a special $V$, let us now consider not the original partition function, but rather a perturbed one, \viz
\eqsc{Z(t)= \int_\mathcal{M} \mathcal{D\phi}~ e^{-S[\Phi]-t \delta V}\label{pertpart}} Note that this function is independent of $t$ as\footnote{recalling $\delta S = 0$ and $\delta^2 V=0.$}
\eqsc{\dfrac{dZ(t)}{dt}=-\int_\mathcal{M} \mathcal{D\phi}~ \delta V e^{-S[\Phi]-t \delta V}=-\int_\mathcal{M} \mathcal{D\phi}~ \delta (V e^{-S[\Phi]-t \delta V})=0 \label{pertpartder}}
This means that the original unperturbed partition function maybe evaluated by evaluating the perturbed partition function $Z(t)$ for any value of $t$ (that is dictated by convenience) and especially, for  $t \rightarrow\infty $. This is immediately useful. If the perturbing operator has a positive definite bosonic part, the integral localizes to a sub-space, often even a finite dimensional one, of field spaces $\left\lbrace \Phi_0\right\rbrace $ where we have $(\delta V)_B \rvert_{\left\lbrace \Phi_0\right\rbrace }=0.$ 

With this motivation, we will try and evaluate the partition function of Supersymmetric CS theory on $L(p,q)$.
Now, to have some supersymmetric actions on some curved 3-manifold, we need to find some background, off-shell supergravity theories that preserve some rigid supersymmetry. These theories can then be made to couple to some supersymmetric field theory. This is done via the stress tensor multiplet. \newline For our specific case of 3d $\mathcal{N}=2$ theory, this supergravity theory was called the ``new minimal supergravity" which has the following field content 
\eqsc{\text{Field Content:~}\big\lbrace \text{Metric}~g_{\mu \nu},~ \text{R Symmetry Gauge Field A}^{(R)}_\mu,~ \text{2-Form Gauge Field B}_{\mu \nu}\nonumber \\\text{Central Charge Symmetry Gauge Field C}_\mu, \text{Gravitini}~ (\psi^\mu, \tilde{\psi}^{\mu})  \big\rbrace }
We define the (dualized) field strengths 
\eqsc{H \equiv \frac{i}{2} \epsilon^{\mu \nu \rho} \partial_\mu B_{\nu \rho } , \qquad V^{\mu} \equiv -i \epsilon^{\mu \nu \rho} \partial_\nu C_{ \rho }}
To ensure that we have rigid supersymmetry, we need to find Killing spinors $(\zeta, \tilde{\zeta})$ which satisfy the Killing spinor equations, given in terms of these fields, as
\eqsc{(\nabla_\mu-i \text{A}^{(R)}_\mu)\zeta = - \frac{1}{2} H  \gamma_\mu \zeta - i V_\mu \zeta - \frac{1}{2} \epsilon_{\mu \nu \rho}V^\nu \gamma^\rho \zeta \nonumber \\ (\nabla_\mu+i \text{A}^{(R)}_\mu)\tilde{\zeta} = - \frac{1}{2} H  \gamma_\mu \tilde{\zeta} + i V_\mu \tilde{\zeta} + \frac{1}{2} \epsilon_{\mu \nu \rho}V^\nu \gamma^\rho \tilde{\zeta} \label{KillingSpinorEqn}} 
In terms of these Killing spinors, the general Supersymmetric variations of the fields in the gauge multiplet for the 3d $\mathcal{N}=2$ theory are given by \footnote{$\mathcal{F}_{\mu \nu} \equiv \partial_\mu \mathcal{A}_\nu - \partial_\nu \mathcal{A}_\mu - i [\mathcal{A}_\mu, \mathcal{A}_\nu] \\ ~~~~~~~D_\mu \equiv \nabla_\mu -i q_R (A_\mu - \frac{1}{2} V_\mu) - i \mathcal{A}_\mu , q_R$  being the R charges of the fields of the vector multiplet . }
\eqsc{\delta \mathcal{A}_\mu = -i (\zeta \gamma_\mu\tilde{\lambda}+ \tilde{\lambda} \gamma_\mu\zeta)~, \nonumber \\\delta \sigma = -\zeta \tilde{\lambda}+ \tilde{\lambda} \zeta~, \nonumber \\ \nonumber \delta \lambda = - \frac{i}{2}\epsilon^{\mu \nu \rho}\gamma_{\rho } \zeta \mathcal{F}_{\mu \nu} +i \zeta (D+\sigma H) - \gamma^\mu \zeta (i D_\mu \sigma - V_\mu \sigma) ~,  \\ \delta \tilde{\lambda} = - \frac{i}{2}\epsilon^{\mu \nu \rho}\gamma_{\rho } \tilde{\zeta} \mathcal{F}_{\mu \nu} -i \tilde{\zeta} (D+\sigma H) + \gamma^\mu \tilde{\zeta} (i D_\mu \sigma + V_\mu \sigma) ~, \nonumber \\  \delta D = D_\mu (\zeta \gamma^\mu \tilde{\lambda} - \tilde{\zeta} \gamma^\mu \lambda ) - i V_\mu (\zeta \gamma^\mu \tilde{\lambda} + \tilde{\zeta} \gamma^\mu \lambda) - [\sigma, \zeta \tilde{\lambda}]- [\sigma, \tilde{\zeta} \lambda ]- H (\zeta \tilde{\lambda}-\tilde{\zeta} \lambda	) \label{SUSYTr} }
We also recall that the 3d $\mathcal{N}=2$ super Yang-Mills (SYM) action on $S^3$, given by\footnote{ Recall $ \mathcal{V}=\left\lbrace  A_\mu,\bar{\lambda},\lambda,\sigma,D\right\rbrace$ is the field content of the 3d $\mathcal{N}=2$ theory. They are respectively a vector, two complex fermions, a scalar and an auxiliary scalar respectively.}\footnote{Actually this is the action given not just on $S^3$ but also on quotient spaces of the kind $S^3/\mathbb{Z}_p$. This is understood as such spaces are locally equivalent to 3-spheres and transformations generated by supercharges are local.}
\eqsc{S_{SYM}=\int d^3x \sqrt{|g|}\Tr\left[ \dfrac{1}{4}\mathcal{F}_{\mu \nu} \mathcal{F}^{\mu \nu}+\dfrac{1}{2} D_\mu \sigma D^\mu \sigma-\dfrac{1}{2}\bigg(D+\dfrac{i}{l}\sigma\bigg)^2-i \bar{\lambda}\gamma^\mu D_\mu \lambda+i\bar{\lambda}[\sigma,\lambda]-\dfrac{1}{2l} \bar{\lambda} \lambda \right] \label{SSYMI} } can also be written as
\eqsc{S_{SYM}=\int d^3x \sqrt{|g|} ~ \dfrac{1}{\tilde{\zeta}\zeta}\delta_{\zeta} \delta_{\tilde{\zeta}} \Tr\left[\dfrac{1}{2}\bar{\lambda}\lambda + i \sigma D \right] \label{SSYMII}}
The action given by Eq. (\ref{SSYMI}) is invariant under the transformations given by Eq.  (\ref{SUSYTr}).\\
Eqs. ($\ref{SSYMI}$) and  ($\ref{SSYMII}$) hand us a prime candidate for the operator $\delta V$ mentioned in the preceding paragraph, \viz, $\delta V = S_{SYM}$. Explicitly, its variation under Grassmann odd symmetry $\delta_{\zeta}$ is zero and has manifestly positive definite bosonic part.\\
So, we would like to evaluate 
\eqsc{Z(t) \equiv \int_\mathcal{M} \mathcal{D\phi}~ e^{-S_{SCS}[\mathcal{V}]-t S_{SYM}} \label{saddle}} with $\mathcal{M}=L(p,q)$ and in the limit $t\rightarrow\infty $ where the partition function localises to a finite dimensional integral and the evaluation is exact. The bosonic part of Eq. ($\ref{SSYMI}$), being expressed as the sum of squares, immediately gives us the BPS configurations. They are \viz,
\eqsc{\mathcal{F}_{\mu \nu}=0 \ , \qquad D_{\mu} \sigma=0 \ , \qquad D+ \frac{i}{l} \ \sigma = 0 \label{8}} 
Here, solving the equations in ($\ref{8}$), we face non trivialities due to difference in global topology of $L(p,q)$ when compared to $S^3$. 

It is evident that we need the classical saddles corresponding to Eq. \eqref{saddle} on $L(p,q)$ on which the localized partition function will be supported. Non-triviality of this statement arises from the fact that the flat connections on a manifold are characterized by holonomies around non-contractible loops on the base manifold, modulo a homogeneous adjoint group action at the base point of the loop. These loops form the first fundamental group of the base manifold. Hence the moduli space of space of flat connections modulo gauge transformation is given by
\bea
\mathrm{hom} \left(\pi_1 (M) \rightarrow G \right) /\mathrm{Ad}_G.
\eea
For the present case, $L(p,q)$ is a free $\mathbb{Z}_p$ quotient of the simply connected manifold $S^3$. Therefore we have the first homotopy group as $\mathbb{\pi}_1$($L(p,q)$)= $\mathbb{Z}_p$. This implies that the CS saddles ie, the flat connections are labelled by $g\in G$, with $g^p$ = 1. If we take $g$ to lie in the maximal torus (this can be always be done for simply connected lie groups by the Ad action), we have 
\eqsc{g=e^{\tfrac{2 \pi i}{p}\mathfrak{m}}, \qquad  \mathfrak{m} \in \Lambda/(p\Lambda) \qquad  \label{9}}
where, $\Lambda$ is the co-weight lattice of the group $G$ and $\mathfrak{m}$ is $N$ dimensional vector, where $N$ is the rank of group $G$. \\
Note that Eq. (\ref{9}) would then imply that $\mathfrak{m}_j\in \mathbb{Z}_p$ . For example, for $G=SU(N)$, we have 
\eqsc{g=\text{diag}\big(~~ e^{\tfrac{2\pi i \mathfrak{m}_1}{p}},~e^{\tfrac{2\pi i \mathfrak{m}_2}{p}},~ ...~ ,~ e^{\tfrac{2\pi i \mathfrak{m}_N}{p}}~~\big) \label{10}} 
with $\sum_{i} \mathfrak{m}_i = 0$. 
The rest of the equations in the Eq. ($\ref{8}$), imply  
\eqsc{\sigma= i l D \equiv \frac{\hat{\sigma}_0}{l} = \text{constant}~, \qquad [\hat{\sigma}_0,\mathfrak{m}]=0 \label{11}} 
We will take $\hat{\sigma}_0$ to lie in the Cartan sub-algebra $\mathfrak{h}$ of the Lie Algebra $\mathfrak{g}$ of the group $G$. Note that,  the second equation of Eq. ($\ref{8}$) motivates why we can expand $\mathfrak{m}$ in the same Cartan basis. \paragraph{Classical Contribution :} The classical (tree level) contribution to the action is obtained by plugging in the BPS configurations in $S_{SCS}$.\nl There will be two such contributions, one coming from the scalars, $\sigma$ and D, which have been shown to be constant in Eq.($\ref{11}$) and the flat gauge fields.  The contribution from the scalars is 
\eqsc{S^I_{SCS}(\hat{\sigma}_0)=\dfrac{i~ \text{vol}(S^3/\mathbb{Z}_p)}{2 \pi l^3}\Tr (\hat{\sigma}^2_0) = \dfrac{\pi i}{p} \Tr(\hat{\sigma}^2_0) \label{12}}
The contribution from the flat gauge fields is 
\eqsc{S^{II}_{CS}(\mathfrak{m}) =- \dfrac{\pi i}{p} \Tr(q^*\mathfrak{m}^2) \label{13}}The total classical contribution is then 
\eqsc{S_{SCS}(\hat{\sigma}_0,\mathfrak{m}) =S^I_{SCS}(\hat{\sigma}_0)+S^{II}_{CS}(\mathfrak{m}) =\dfrac{\pi i}{p} \Tr(\hat{\sigma}^2_0-q^* \mathfrak{m}^2) \label{14}} with $q^*$ is defined as $q^* q = 1~ \text{mod($p$)}$ \paragraph{1-Loop Determinants :} We calculate the 1-Loop determinants from the quadratic fluctuations of the fields about their BPS configurations. Specifically,
\eqsc{A_\mu = t^{-\frac{1}{2}} A^{\prime}_\mu~,~\sigma = \frac{\hat{\sigma}_0}{l}+ t^{-\frac{1}{2}} \sigma^{\prime}~,~D=- \frac{i}{l^2}\hat{\sigma}_0 + t^{-\frac{1}{2}} D^{\prime}~,~\lambda= t^{-\frac{1}{2}} \lambda^{\prime}~,~\bar{\lambda}= t^{-\frac{1}{2}} \bar{\lambda}^{\prime} \label{quadfluc}}\\
Plugging these values in Eq.(\ref{SSYMI}), we obtain the terms in the action proportional to t$^{-1}$ as 
\eqsc{S^{\prime}_{SYM}=t^{-1} \int d^3x \sqrt{|g|}\Tr\bigg[ \dfrac{1}{4}F^{\prime}_{\mu \nu} F^{\prime \mu \nu}+\dfrac{1}{2} \partial_\mu \sigma^{\prime} \partial^\mu \sigma^{\prime}-\dfrac{1}{2l^2} [A^{\prime}_\mu,\hat{\sigma}_0]^2 \nonumber \\ - \dfrac{1}{2}\bigg(D^{\prime}+\dfrac{i}{l}\sigma^{\prime}\bigg)^2 -i \bar{\lambda}^{\prime}\gamma^\mu D_\mu \lambda^{\prime}+\frac{i}{l}\bar{\lambda}[\hat{\sigma}_0^{\prime},\lambda^{\prime}]-\dfrac{1}{2l} \bar{\lambda^{\prime}} \lambda^{\prime} \bigg] + \mathcal{O}(t^{-2})\label{SprimeSYM}}\\
The integration over $D^\prime$ can trivially be done and it just alters the overall normalization constant sitting in front. 
To deal with the Vector and Fermionic fields, we decompose the gauge field into a divergenceless part $(X)$ and the rest as
\eqsc{A^{\prime}_\mu = X_\mu + \partial_\mu \phi} 
The integrals over $\phi$ and $\sigma^\prime$ give determinants that cancel and we are left with a divergenceless Vector field and Fermionic fields. Next, we expand them in the $\Gamma_a$ of the Lie Algebra with the definition 
\eqsc{[\hat{\sigma}_0,\Gamma_\alpha] = \alpha(\hat{\sigma}_0)}
The action then becomes 
\eqsc{\int  d^3x \sqrt{|g|}	\sum_{\alpha \in Ad(G)} \bigg(\frac{1}{2}X^\mu_{-\alpha} \big(-\nabla^2+\frac{1}{l^2}\alpha(\hat{\sigma}_0)^2\big)X_{\alpha,\mu}+\tilde{\lambda}^{\prime}_{-\alpha}\big(-i \gamma^\mu \nabla_\mu + \frac{i}{l} \alpha(\hat{\sigma_0}) - \frac{1}{2l}\big)\lambda^{\prime}_{\alpha}\bigg) }
The 1-Loop Determinant is then, simply 
\eqsc{\mathcal{Z}_{gauge}^{1-Loop} (\hat{\sigma}_0,\mathfrak{m};p,q) = \dfrac{\text{det}\big(-i \gamma^\mu \nabla_\mu + \frac{i}{l} \alpha(\hat{\sigma_0}) - \frac{1}{2l}\big)}{\text{det}(-\nabla^2+\frac{1}{l^2}\alpha(\hat{\sigma}_0)^2\big)^\frac{1}{2}} }\\
On Lens Spaces $L(p,q)$, this result may be calculated as :
\eqsc{\mathcal{Z}^{1-Loop}_{gauge}(\hat{\sigma}_0,\mathfrak{m};p,q) = \dfrac{4~\prod_{\alpha >0} \sinh \dfrac{\pi}{p} \alpha(\hat{\sigma}_0 + i \mathfrak{m} )\sinh \dfrac{\pi}{p} \alpha(\hat{\sigma}_0 - i q^* \mathfrak{m} ) }{\prod_{\alpha>0}   \alpha(\hat{\sigma_0})^2} \label{15}} where, $\alpha$ are the roots of $G$ and $q^*$ is defined as $q^* q = 1~ \text{mod($p$)}$.

For a detailed derivation of the result in Eq.$(\ref{15})$, we refer the reader to \cite{Gang:2009wy} \footnote{Note, that supersymmetric CS theory admits Matter Multiplets (MM) too, in arbitrary representation $\mathfrak{R}_i$ for the i-th multiplet, and indeed, in the literature, the full theory has been localized, but, for our purposes, we would not require any MM. }. We only draw our reader's attention to the fact that the above expression reduces to the 1-Loop determinant of the partition function evaluated on $S^3$ for the special case of $p=1$ and $q^*=0$ as it should as $L(1,0)=S^3$. \\
Finally, we integrate over the BPS configurations, here, denoted by $\sigma_i$'s and sum over the holonomies, identified by the components of the vector $\mathfrak{m}$. Using Weyl Integration formula, as always, the integral reduces from the vector space spanned by the entire Lie Algebra to a  vector subspace spanned by just the Cartan Sub-Algebra ($\mathfrak{h}$). This, however, introduces a Vandermonde Determinant $\prod_{\alpha>0} \alpha(\hat{\sigma_0})^2$. This is exactly cancelled by the denominator in Eq. $(\ref{15})$. Also, to take into account the residual Weyl symmetry of the gauge group, we divide the final expression by the order of the Weyl group of the the gauge group. \nl
Explicitly, the expression for the partition function becomes 
\eqsc{\mathcal{Z}(\hat{\sigma}_0,\mathfrak{m};p,q)=\dfrac{1}{|\mathcal{W}|}\sum_{\mathfrak{m}} \int_{\mathfrak{h}} d\hat{\sigma}_0 ~\prod_{\alpha>0} \alpha(\hat{\sigma_0})^2~e^{-S_{SCS}}~\mathcal{Z}^{1-Loop}_{gauge}(\hat{\sigma}_0,\mathfrak{m};p,q) \nonumber \\ =\dfrac{4}{|\mathcal{W}|}\sum_{\mathfrak{m}} \int_{\mathfrak{h}}d\hat{\sigma}_0~~e^{- \tfrac{\pi i}{p} \Tr_{\text{CS}}(\hat{\sigma}^2_0-q \mathfrak{m}^2) }  \prod_{\alpha >0} \sinh \dfrac{\pi}{p} \alpha(\hat{\sigma}_0 + i \mathfrak{m} )\sinh \dfrac{\pi}{p} \alpha(\hat{\sigma}_0 - i q^* \mathfrak{m} )  \label{16}} 
We will evaluate the RHS of Eq.  ($\ref{16}$) explicitly next. 
\subsection{Partition Function : Evaluation of the Matrix Model for spin-2 Gravity} \label{partfunc}
As described in the section \ref{thetheory}, the CS version of the spin-2 gravity we are interested in is based on the gauge group $G=SU(2) \times SU(2)$ for the gauge fields in Eq. Eq. \eqref{mod}. Here we would perform the localized integral Eq. \eqref{16} and choose those CS saddles that correspond to smooth gravity background solutions. 

Let us, then, evaluate the partition function given by Eq.  ($\ref{16}$) for $G=SU(2)\times SU(2)$. \nl 
The Weyl Group for $SU(N)$ is the permutation group S$_N$ , the order of which is $N!$. The rank of $SU(N)$ group is $(N-1)$, which, for our case of $SU(2)$ is simply 1. Hence, flat connections are identified by the component of a one component vector $\mathfrak{m}$, denoted by $m_{\pm}$ for the two gauge fields $\mathcal{A}_{\pm}$ corresponding to the two $SU(2)$ groups of the gauge group $G$.

The partition function, for each saddle, identified by a value of p, receive contribution from two values of $\text{m}_\pm$. They are explicitly,
\eqsc{\text{m}_\pm= \dfrac{(q \pm 1)}{2} \label{chosen_saddle}}
For further details, we refer the reader to \cite{Castro:2011xb}.

With the values of m$_\pm$'s in our hand, we can directly proceed to calculate the integral given in the RHS of Eq.  ($\ref{16}$) explicitly
\nl As discussed, since the rank of $SU(2)$ group is 1, the evaluation of the partition function reduces to the problem of solving a one dimensional integral, \viz :
\eqsc{\mathcal{Z}_+(\hat{\sigma}_0,\mathfrak{m};p,q)=\dfrac{4}{2!} \int d\lambda_+~~e^{- \tfrac{i k_+\pi }{p} (\lambda_+^2-q^* \text{m}_+^2) }  \sinh \dfrac{\pi}{p} (\lambda_+ + i \text{m}_+ )\sinh \dfrac{\pi}{p} (\lambda_+ - i q^* \text{m}_+)  \label{17}}
Fortunately, the integral given in Eq.  ($\ref{17}$) is tractable.\nl
Since our chosen gauge group is a product group we have another flat connection, identified by m$_-$. The corresponding CS level is denoted by $k_-$ and we obtain an equivalent expression for the second flat connection. Explicitly,
\eqsc{\mathcal{Z}_-(\hat{\sigma}_0,\mathfrak{m};p,q)=-\dfrac{4}{2!} \int d\lambda_-~~e^{- \tfrac{i k_-\pi }{p} (\lambda_-^2-q^* \text{m}_-^2) }  \sinh \dfrac{\pi}{p} (\lambda_- + i \text{m}_- )\sinh \dfrac{\pi}{p} (\lambda_- - i q^* \text{m}_-)  \label{18}}
As yet, the CS levels are arbitrary but we will choose a special parametrization, \viz ,
\eqs{k_+=a(\dfrac{1}{\gamma}+1)~,\qquad k_-=a(\dfrac{1}{\gamma}-1) \label{19}} Here, $\gamma$ is a tunable parameter, whose large limit, for \eg , reproduces $k_+ + k_- = 0$ . However, we would focus on the small $\gamma$ regime for the purpose of divergence structure of the partition function.  \nl The total contribution to the partition function is their product. Explicitly, \eqsc{\underbrace{\mathcal{Z}(\hat{\sigma}_0,\mathfrak{m};p,q)}_{\mathfrak{su}(2)\oplus \mathfrak{su}(2)}=\underbrace{\mathcal{Z}_+(\hat{\sigma}_0,\mathfrak{m};p,q)}_{\mathfrak{su}(2)} \times \underbrace{\mathcal{Z}_-(\hat{\sigma}_0,\mathfrak{m};p,q)}_{\mathfrak{su}(2)}\label{20}} 
Using the the values of  m$_+$ and m$_-$, the RHS of Eq.  ($\ref{20}$) gives
\eqsc{\mathcal{Z}(\hat{\sigma}_0,\mathfrak{m};p,q)=\frac{i p \gamma}{(2!)^2a\sqrt{1-\gamma^2}}e^{\tfrac{i \pi (a (q+q^*+2\gamma)-4(1+q)\gamma)}{2p\gamma}}\bigg(1+e^{\tfrac{4i \pi (1+q)}{p}}+e^{\tfrac{2 i \pi (q-q^*)}{p}}+\nonumber \\e^{\tfrac{2i \pi (2+q+q^*)}{p}}-e^{\tfrac{2i \pi (a(1-q)(1-\gamma)+2\gamma)}{ap(\gamma-1)}} - e^{\tfrac{2i \pi (a(3+q)(-1+\gamma)+2\gamma)}{ap(\gamma-1)}}-e^{\tfrac{2i \pi (a(q^*-1)(1-\gamma)+2\gamma)}{ap(\gamma-1)}}-\nonumber \\ 2e^{\tfrac{2i \pi (a(1+q)(1+\gamma)-2\gamma)}{ap(\gamma+1)}}-e^{\tfrac{2i \pi (a(1+2q-q^*)(1+\gamma)-2\gamma)}{ap(\gamma+1)}} -e^{\tfrac{2i \pi (a(1+q^*)(1+\gamma)-2\gamma)}{ap(\gamma+1)}}+e^{\tfrac{4i \pi (a(\gamma^2-1)+2\gamma)}{ap(1+\gamma)(1-\gamma)}}+\nonumber \\e^{\tfrac{4i \pi (aq(\gamma^2-1)+2\gamma)}{ap(\gamma+1)(\gamma-1)}}+e^{\tfrac{2i \pi (a(2+q-q^*)(\gamma^2-1)+4\gamma)}{ap(\gamma+1)(\gamma-1)}}+e^{\tfrac{2i \pi (a(q+q^*)(\gamma^2-1)+4\gamma)}{ap(\gamma+1)(\gamma-1)}}-e^{\tfrac{2i \pi (a(1+2q+q^*)(\gamma-1)+2\gamma}{ap(\gamma-1)}} \bigg)\label{21}}
This is one of the most striking points in our analysis, which requires further attention. We should note that, the above expression is same as that appearing in the purely bosonic analysis of \cite{Basu:2011vs} or the one in \cite{Castro:2011xb} (for $\gamma \rightarrow \infty$), apart from the overall pre-factor $p$.
For this purpose we take $\gamma \rightarrow \infty$ and large $a$ in \eqref{21} with an analytical continuations $a \rightarrow i \, a$. For large $|a|$ (which means assuming large dS radius in comparison to Planck length), ie where we expect the CS theory to be describe quantum gravity, \eqref{21} yields:
\bea \label{compare_ours}
\mathcal{Z}(a, p, q) = \frac{8 \pi^2}{a\, V_{L(p,q)}} F(a,q,p).
\eea
Whereas the result for pure bosonic gravity, as found in \cite{Castro:2011xb} \footnote{There lingers a typo in \cite{Castro:2011xb}, particularly in eq. (4.32), involving extra factors of $2$ in the cosine phases}, which also is a special case for higher spin result of \cite{Basu:2015exa} is:
\bea \label{compare_castro}
&& \mathcal{Z'}(a, p, q) =\frac{V_{L(p,q)}}{4 a \pi^2} \, F(a,q,p), \\
&& \mbox{where }\, F(a,q,p) = e^{\tfrac{2 \pi k}{p}} \left(\left(\cos(\tfrac{\pi}{p})- \cos(\tfrac{\pi q}{p})\right)\left(\cos(\tfrac{\pi}{p})- \cos(\tfrac{\pi q^*}{p})\right)\right). \nonumber
\eea
Here, $V_{L(p,q)} = 2 \pi^2/ p$ is the volume of $L(p,q)$, measured in units of dS length cubed. This clearly shows that inclusion of fermionic modes basically introduced an altered power law volume dependence. This factor, as explained also in the introductions, would cancel if one is interested in local physics ie.  calculate correlation functions on a particular Lens space. However as already motivated, we defer that analysis here and go on finding an answer to a question rather topological in nature.
We sum over all poassible gravity saddles, ie. Lens spaces. In short, the overall contribution to the gravity partition function $\mathcal{Z}_{\text{gravity}}$, we will have a sum over $p$ and sum over $q$ to accommodate the various contributions of all the saddles. In short, the gravity partition function will be obtained by : 
\eqsc{\mathcal{Z}_{\text{gravity}}=\sum_{p=1}^{\infty} \sum_{\substack{q=1 \\(p,q)=1}}^p\mathcal{Z}(\hat{\sigma}_0,\mathfrak{m};p,q).\label{22}}
We observe an overall positive power of $p$ multiplying the trigonometric terms. When summed over all topologies, ie. lens spaces, this $p$ dependence might be a serious cause of divergence. Interestingly, for the pure bosonic theory (for $ \gamma \rightarrow \infty$) \cite{Castro:2011xb} and (finite $\gamma$) \cite{Basu:2011vs} the overall $p$ dependence was $1/p$. Therefore the  supersymmetric theory  does not reproduce exactly the same answer as that of the purely bosonic theory . We will shortly come back to the detailed analytic structure of the sum and explore deeper in this aspect.
We will express our result, after the sum over $q$'s in terms of Kloosterman Sums $S(x,y;p$), which are tailor made for such sums. The Kloosterman sums are defined as  
\eqscn{S(x,y;p)\equiv \sum_{\substack{q=1 \\(p,q)=1}}^p e^{\frac{2i \pi}{p}(xq + yq^*)}} 
In terms of these Kloosterman sums, the $q$ sum in Eq.  ($\ref{22}$) gives\footnote{$ \alpha \equiv \dfrac{a}{4  \gamma}$} :
\eqsc{\mathcal{Z}_{\text{gravity}}=\frac{i}{(2!)^2}\sum_{p=1}^{\infty}\dfrac{ p \gamma}{ \ a \sqrt{1-\gamma^2}} \ e^{\tfrac{i \pi a}{p}}\bigg[ 4 \cos\big(\frac{2\pi}{p}\big)\bigg(S(\alpha -1,\alpha ;p)+S(\alpha +1,\alpha ;p)\bigg)- \nonumber \\ 2\big( \cos\big(\frac{4 \pi}{p}\big)+1\big)S(\alpha ,\alpha ;p) - \nonumber \\ \bigg(S(\alpha-1 ,\alpha -1;p)+2S(\alpha+1 ,\alpha -1;p)+S(\alpha+1 ,\alpha +1;p)\bigg)\bigg] \label{23}} 
To carry out the summation over $p$, we expand the cosine and the exponential function in their respective Maclaurin series. We obtain an infinite series of Kloosterman Zeta function, defined as 
\eqsc{L(x,y;s)=\sum_{p=1}^{\infty}p^{-2s}S(x,y;p)\label{24}} The Kloosterman Zeta functions are analytic for $\mathfrak{Re}(s)>1/2$. Writing our result explicitly, in terms of these functions, will also help us isolate the divergent pieces in the gravity partition function, as explicitly those terms with $\mathfrak{Re}(s)\leq1/2$ . The final expression for $\mathcal{Z}_{\text{gravity}}$ is then obtained as :
\eqsc{\mathcal{Z}_{\text{gravity}}=\dfrac{i  \ \gamma}{(2!)^2a \sqrt{1-\gamma^2}}\sum_{m=0}^{\infty} \dfrac{(i \pi a)^m}{m!}\bigg[ \sum_{n=0}^{\infty} (-1)^n \dfrac{ 4 \ \pi^{2n}}{(2n)!} \bigg( L\big(\alpha - \frac{1}{2},\alpha,\dfrac{m+2n-1}{2}\big) + \nonumber \\ L\big(\alpha + \frac{1}{2},\alpha,\dfrac{m+2n-1}{2}\big) -2^{2n-1} L\big(\alpha ,\alpha,\dfrac{m+2n-1}{2}\big) \bigg) - 2 L\big(\alpha ,\alpha,\dfrac{m-1}{2}\big) -  \nonumber \\L\big(\alpha - \frac{1}{2},\alpha,\dfrac{m-1}{2}\big)- 2 L\big(\alpha - \frac{1}{2},\alpha + \frac{1}{2},\dfrac{m-1}{2}\big)- L\big(\alpha ,\alpha - \frac{1}{2},\dfrac{m-1}{2}\big) \bigg] \label{24A}} 
Let us investigate the analytic structure of the partition function summed over all Lens spaces Eq. \eqref{24A}. From Eq. \eqref{24}, ie the analyticity of the Kloosterman zeta function, it is easy to see a set of divergence is sourced from the terms for which $m+2n\leq2$ in Eq. \eqref{24A} and another set being originated from $m \leq 2$ for $n$ independent terms.

It might be a bit more instructive to review the divergence properties in semi-classical regime along with $\gamma \rightarrow \infty$, so that a direct comparison with the $\zeta(1)$ divergence appearing in \cite{Castro:2011xb} can be made. This is actually made apparent by comparing \eqref{compare_castro} and \eqref{compare_ours}. Even in the milder case of purely bosonic theory, which depends linearly on volume as an over-all factor, a divergence occurs when one considers sum over all Lens spaces as a harmonic series in $p$, since $V_{L(p,q)} \sim 1/p$, ie very slowly while accumulating up smaller Lens space volumes. However, we should keep in mind that, this divergence is completely different in nature to those commonly seen in local QFTs while probing short length-scales, ie. the UV divergences. Those originate from integrating high energy modes. For a renormalizable theory these divergences can be absorbed into local counter-terms. We don't have any such mechanism here, as also commented in \cite{Castro:2011xb}.

In contrast, our analysis shows existence of more such divergent pieces in \eqref{24A}, due to dynamical fermions due to localization. Individual Lens space contributions are finite as before but summing over Lens spaces makes the divergence worse. As the prior motivation for this sum over saddles was to inspect the Hartle-Hawking state for the static patch of de Sitter space, the present result summarizes that quantum Hartle-Hawking is not a good choice of vacuum for 3 dimensional dS, even in the supersymmetrized version.
\section{Higher Spin Case }  \label{hahah}
Linearized higher spin fields can be coupled consistently to gravity in 3 dimensions with finite height of the higher spin tower, which is nicely captured by the Fronsdal action of symmetric traceless tensor fields. In principle, we imagine a (finite) tower of higher spins, namely s = 3, 4, 5, ... , N over and above the spin-2 cases. This construction is possible only in three dimensions where we can have a consistent truncation to arbitrary spins. For higher dimensions $(d>3)$ we must include all the higher spin fields. In three dimensions, however, we have the added advantage where we can have a truncated tower of higher spin fields. 

These higher  spin fields are all minimally coupled to the spin-2 field which forms a background. Following the analysis put forward in \cite{Gaberdiel:2010ar}, we would include higher spins in our analysis and evaluate the partition function and see the nature of divergence, if any. We would explicitly work out the effect of adding a spin-3 fields as that is the most tractable case in these theories of higher finite spins. As explained, the background is still furnished by the spin-2 field such that $g_{\mu \nu}$ remains the metric of the static patch of Euclidean de Sitter spacetime, given by (\ref{met1}). We define a metric compatible connection $\nabla$ on the manifold such that
\begin{eqnarray}
\left[ \nabla_{\mu},\nabla_{\nu} \right] A^{\rho}= R^{\rho}_{~\sigma \mu \nu} A^{\sigma}
\end{eqnarray} which defines the Riemann Curvature tensor on the manifold for a probe field $A^{\mu}$.
\paragraph{Spin-3 case} To introduce a massless spin-3 field $\phi_{\left( \mu \nu \rho\right) }$ which is minimally coupled to pure gravity in 3 dimensions, we introduce, following \cite{Fronsdal:1978rb}, the linearised Fronsdal action given by 
\begin{eqnarray}
S[\phi]= \int d^3 x \sqrt{g}~ \phi^{\alpha_1 \alpha_2 \alpha_3} \left(\mathcal{H}_{\alpha_1 \alpha_2 \alpha_3} -\frac{1}{2}g_{(\alpha_1 \alpha_2} \mathcal{H}_{\alpha_3) \mu }^{~~~~~\mu} \right) 
\end{eqnarray}
where the definitions are as follows,
\begin{eqnarray}
	\mathcal{H}_{\alpha_1 \alpha_2 \alpha_3} \equiv \Delta \phi _{\alpha_1 \alpha_2 \alpha_3} - \nabla_{(\alpha_1} \nabla ^\lambda \phi_{\alpha_2 \alpha_3) \lambda} + \frac{1}{2} \nabla_{(\alpha_1} \nabla_{\alpha_2} \phi_{\alpha_3)\lambda}^{~~~~\lambda} +2g_{(\alpha_1 \alpha_2} \phi_{\alpha_3) \lambda}^{~~~~\lambda} 
\end{eqnarray} 
We also note, in passing, that the linearised theory enjoys a gauge symmetry given by
\begin{eqnarray*}
\delta \phi_{\alpha_1 \alpha_2 \alpha_3 } = \nabla_{(\alpha_1} \xi_{\alpha_2 \alpha_3)}
\end{eqnarray*} where, $\xi_{\alpha \beta}$ is symmetric and traceless.

Interestingly, a first order version of this theory can also be formulated in terms of CS gauge fields. \cite{Campoleoni:2010zq} gives an elaborate AdS counterpart of that exposition. Our work is similar in spirit but with a positive cosmological constant, which, has not been explored before. At the level of corresponding Lie algebra for CS theory, going from AdS to dS background amounts to changing the structure constants. The CS theory that describes spin 3 fields on the backdround of (euclideanised) 3d dS spacetime, has a gauge group $SU(3) \times SU(3)$ \cite{Basu:2015exa}. 

For the ease of generalizing to spin-3 case, in spirit of the Eq.($\ref{gaugeflddef}$), we define
\begin{eqnarray}
(j_{\pm})_\mu ^{~p} \equiv  (\omega \pm e)_\mu^{~p} \label{potential}
\end{eqnarray} 
Let us further define higher tensorial objects obtained similarly as a linear combinations gauge potentials
\begin{eqnarray}
(t_{\pm})_{\mu}^{~p_1 p_2 ... p_{s-1}} \equiv (\omega \pm e)_\mu^{~p_1 p_2 ... p_{s-1}}\label{highspinpotential}
\end{eqnarray}
We then define the one form gauge fields as
\begin{eqnarray}
\mathcal{A}^{\pm} \equiv ((j_{\pm})_\mu ^{~p} J_p + (t_\pm)_\mu^{~p_1 p_2 ... p_{s-1}} T_{p_1 p_2 ... p_{s-1}}) dx^\mu \label{highergaugefield}
\end{eqnarray} where $\left\lbrace T_{p_1 p_2 ... p_{s-1}} \right\rbrace $ are higher spin generators which are to be added to  $\left\lbrace j_p \right\rbrace $.

Here too, there are no local degrees of freedom, and we associate the equations of motion with the condition for flatness for these gauge fields. This is, again, similar in spirit to the $d=3$ Einstein gravity we explored earlier. Thus, we arrive at the Chern Simons formulation of higher spin gravity. 

Explicitly, we need to state the algebra of these higher spin generators $\left\lbrace T_{p_1 p_2 ... p_{s-1}} \right\rbrace $. Firstly, we note that, from Eqs. (\ref{potential}), (\ref{highspinpotential}) \& (\ref{highergaugefield}), the generators must transform in some irreducible representation of $su(2)$. This implies that they are symmetric and traceless. Furthermore, similar to the $\left\lbrace J_p \right\rbrace $, they satisfy
\begin{eqnarray}
\nonumber \left[ J_q,J_r \right] &=& \epsilon_{qrt}J^t \\
\left[ J_r,T_{~p_1 p_2 ... p_{s-1}} \right]&=& \epsilon^q_{~~r(p_1}T_{p_2 p_3 p_{s-1})q}  \label{genalgebra}
\end{eqnarray}
Particularly, for the case of $s=3$, Eq. (\ref{genalgebra}) allows for a non-trivial algebra of the higher spin generators, namely,
\begin{eqnarray}
\nonumber \left[ J_q,J_r \right] &=& \epsilon_{qrt}J^t \\
\nonumber \left[ J_r,T_{p_1 p_2} \right]&=& \epsilon^q_{~~r(p_1}T_{p_2)q} \\
\left[T_{p_1 p_2},T_{p_3 p_4} \right]&=& \left(\delta_{p_1(p_3}\epsilon_{p_4) p_2 r}+\delta_{p_2(p_3}\epsilon_{p_4) p_1 r} \right) J^r \label{s3genalgebra}
\end{eqnarray}
 One can further show that the algebra given by Eq. (\ref{s3genalgebra}) is isomorphic to $su(3)$. That we are working on a Riemannian manifold is made explicit by the appearance of the Kronecker delta as opposed to the Minkowski metric in the algebra Eq. $(\ref{s3genalgebra})$ .

With the set of generators  $\left\lbrace J_p,T_{p_1 p_2 ... p_{s-1}} \right\rbrace $ which generate a Lie Algebra $\mathfrak{g}$, assumed to admit a non-degenerate bilinear form $\Tr$, we define a Chern Simons action 
\begin{eqnarray}
\label{higherspincsaction}
S[\mathcal{A}^+ ,\mathcal{A}^-] &=& \frac{k}{4 \pi} \Tr\int _M ( \mathcal{A}^+ \wedge d\mathcal{A}^+ + \frac{2}{3} \mathcal{A}^+ \wedge\mathcal{A}^+ \wedge \mathcal{A}^+)\nonumber \\&-&\frac{k}{4 \pi} \Tr \int _M ( \mathcal{A}^- \wedge d\mathcal{A}^- + \frac{2}{3}\mathcal{A}^- \wedge\mathcal{A}^- \wedge \mathcal{A}^-).
\end{eqnarray}
We would like to calculate the exact partition function in this case so as to check whether supersymmetric version of the higher spins make the sum over topologies better in terms of convergence properties. Let us now evaluate the partition function given by Eq. ($\ref{16}$) for $G=SU(3)\times SU(3)$. As the rank of the group $SU(3)$ is 2, the flat connections are identified by the component of a two component vector $\mathfrak{m}$, denoted by m$^{(\text{i})}_{\pm}$, where, i running from 1 to 2, denotes the two components of $\mathfrak{m}$ and $\pm$, as before, denote the two gauge fields $\mathcal{A}_{\pm}$ corresponding to the two $SU(3)$ groups of the gauge group $G$.

At this point, as in the case for spin-2 in Eq. \eqref{chosen_saddle}, we will have to choose a pair of elements from the corresponding $A{}_2$ co-weight lattice. This choice is physically motivated by the fact that quantum fluctuations are considered over the background that describes dS geometry in terms of gravitons and zero excitations for higher spin degrees of freedom. The exact co-weight points are thus found by a principal embedding of $\mathfrak{su}(2)$ in $\mathfrak{su}(3)$. 
Thus the two components of $\mathfrak{m}_{\pm}$ as 
\eqsc{\text{m}^{(i)}_+= \left\lbrace q+1,0 \right\rbrace, \qquad \text{m}^{(i)}_-= \left\lbrace q-1,0 \right\rbrace \label{chosen_saddle_SU(3)}}
With the values of m$^{(i)}_{\pm}$ 's in our hand, we can directly proceed to calculate the integral given in the RHS of Eq.  ($\ref{16}$) explicitly.
\eqsc{\mathcal{Z}_{\pm}(\hat{\sigma}_0,\mathfrak{m};p,q)=\nonumber \\\pm \dfrac{4}{3!} \int d\lambda^{(1)}_{\pm}d\lambda^{(2)}_{\pm}~~e^{- \tfrac{i k_{\pm}\pi }{p} (\lambda^{(1)2}_{\pm}+\lambda^{(2)2}_{\pm}-q^* (\text{m}_{\pm}^{(1)2}+\text{m}_{\pm}^{(2)2})) }  \sinh \dfrac{\pi}{p} (2\lambda^{(1)}_{\pm} - \lambda^{(2)}_{\pm} + i (2\text{m}^{(1)}_{\pm} -\text{m}^{(2)}_{\pm}  )) \times \nonumber \\ \sinh \dfrac{\pi}{p} (2\lambda^{(1)}_{\pm} - \lambda^{(2)}_{\pm} - i q^* (2\text{m}^{(1)}_{\pm} -\text{m}^{(2)}_{\pm}  )) \sinh \dfrac{\pi}{p} (2\lambda^{(2)}_{\pm} - \lambda^{(1)}_{\pm} +i (\text{m}^{(2)}_{\pm} -\text{m}^{(1)}_{\pm}  )) \times \nonumber \\ \sinh \dfrac{\pi}{p} (2\lambda^{(2)}_{\pm} - \lambda^{(1)}_{\pm} - i q^* (\text{m}^{(2)}_{\pm} -\text{m}^{(1)}_{\pm}  )) \sinh \dfrac{\pi}{p} (\lambda^{(1)}_{\pm} + \lambda^{(2)}_{\pm} + i (\text{m}^{(1)}_{\pm} +\text{m}^{(2)}_{\pm}  ))  \times \nonumber \\ \sinh \dfrac{\pi}{p} (\lambda^{(1)}_{\pm} + \lambda^{(2)}_{\pm} - i q^* (\text{m}^{(1)}_{\pm} +\text{m}^{(2)}_{\pm}  )) \label{25}}
The integral in Eq.  ($\ref{25}$) is Gaussian and therefore, tractable. The argument preceding Eq.  ($\ref{20}$) holds in this case too, and we have 
\eqsc{\underbrace{\mathcal{Z}(\hat{\sigma}_0,\mathfrak{m};p,q)}_{\mathfrak{su}(3)\oplus \mathfrak{su}(3)}=\underbrace{\mathcal{Z}_+(\hat{\sigma}_0,\mathfrak{m};p,q)}_{\mathfrak{su}(3)} \times \underbrace{\mathcal{Z}_-(\hat{\sigma}_0,\mathfrak{m};p,q)}_{\mathfrak{su}(3)}\label{27},}
with $k_\pm$ being parameterized similarly as in the $SU(2)$ case, via Eq.  ($\ref{19}$), and the values of m$_{\pm}$ obtained in Eq. \eqref{chosen_saddle_SU(3)}, obtained in the preceding section, the RHS of Eq.  ($\ref{27}$) gives 

\eqsc{\mathcal{Z}(\hat{\sigma}_0,\mathfrak{m};p,q)=\frac{ ( p \gamma)^2}{(3!)^2 a^2(\gamma^2 -1)}~e^{\tfrac{2 i \pi}{p \gamma }(a(q+q^*+2\gamma)-4(1+q)\gamma)} \, G(\gamma, a , p, q).
\label{28}}
Here the function $G$ is a linear combination of 824 phase factors, similar in form, to those appearing inside \eqref{21}. Due to the cumbersome appearance and of less significance of these terms, they have been omitted here.

Again, following similar arguments as before, the gravity partition function is given by a sum over the topologies, which classify the various saddles, and is obtained as 
\eqsc{\mathcal{Z}_{\text{gravity}}=\sum_{p=1}^{\infty} \sum_{\substack{q=1 \\(p,q)=1}}^{p}\mathcal{Z}(\hat{\sigma}_0,\mathfrak{m};p,q)\label{29}}
Even without knowing the explicit structure of the terms in the right hand side of Eq. \eqref{28}, just from the pre-factor $p^2$ we can conclude as in the spin-2 case that Eq. \eqref{29} will diverge because of terms appearing in the non-analytic domain of Kloosterman zeta function.

We conclude by a comparative remark with the purely bosonic theory. 
For example, \cite{Basu:2015exa} states, that the partition function for a purely bosonic theory of higher spins truncated at a tower of spin $N$ on a Lens space is given by:
\bea \label{full}
\sim( V_{L(p,q)})^{N-1 }\, e^{2\pi k/p} \prod_{\pm} \prod_{s=2}^{N-1} \prod_{r=1}^{s-1} \sin \left( r \pi  \frac{q \pm 1}{p} \right)\sin \left( r \pi  \frac{q^{ \ast } \pm 1}{p} \right).  
\eea
This makes the sum over topologies more convergent for higher spins.
In the supersymmetrized version however:
$$\mathcal{Z}_{\mathrm{spin-}N} \sim p^{N-1} \sim \frac{1}{V_{L(p,q)})^{N-1 }}.$$  Due to the opposite statistics of the fermions and opposite power of fermionic determinants in partition function calculations, the divergence in the full partition function gets only worse. However, as already discussed several times, this divergence only tells about stability of quantum gravity fluctuations on de Sitter background, but goes away while calculating correlators of non-gravitational interactions on fixed Lens space backgrounds.

\section{Conclusions and future directions } \label{conclusions}
\quad To conclude, let us recall what we have achieved. 
We have calculated, as the definition of Hartle-Hawking vacuum state the exact quantum gravity partition function on the static patch of Euclidean de Sitter space-time. In trying to do so, we have argued that the quantum gravity path integral receives contributions from all the classical saddles, which we have obtained as the quotient spaces of $S^3$ by the abelian group, $\mathbb{Z}_p$. This have been identified with the Lens Space $L(p,q)$ and we expect a formal sum over $p$ and $q$, the parameters of the space to capture the contributions from the saddles.

To evaluate the quantum gravity partition function exactly, we have worked in the CS formulation of 3d gravity. This has proved immediately helpful in calculating the exact partition function by supersymmetric localization technique. We have calculated the partition function for both spin-2 gravity and higher spin cases. We observe that the Kloosterman zeta functions arise naturally in the result of the partition functions from where we identify explicitly the divergent pieces. 
We also observe that our result, being exact, reproduces the known result in large $k$ limit, apart from an overall volume factor. That contribution has been ascribed to the effect of introduction of dynamical fermionic degrees of freedom. Due to the presence of this change in the prefactor, the analytic properties of the sum over all Lens space does change. It becomes divergent even for those ranges of parameters, for which the bosonic theory was finite. This has a serious implication while interpreting the result as a Hartle-Hawking vacuum wave-function.

To explore further, let us focus that the divergence is caused basically from the prefactor volume contributions from Lens space of higher $p$. As one goes on incorporating higher values of $p$, smaller volumes contribute as $p/(2 \pi^2)$ as per \eqref{compare_ours}. 
Therefore one of the most natural yet brute-force regularization would be to consider only those Lens space whose volumes are greater than some particular volume $V_{\Lambda}$, similar in spirit to putting a UV cut-off. One obvious choice for $V_{\Lambda} $ is of course the Planck volume. This gives a seemingly plausible regulator. However the ultimate physicality of this scheme would be to first compute expectation value of local operators or correlators and then take $V_{\Lambda} \rightarrow 0$ and check that the results converge uniformly to a finite limiting value. That would make a very clear sense of the Hartle-Hawking vacuum for 3 dimensional dS space, with all quantum gravity effects included. In fact the above scheme is planned for an immediate future check, which we would like to perform by including local degrees of freedom in the form of scalar fields. The present results of this article from the localized gravitational part of the path integral would make that calculation relatively more tractable.

For the higher spin cases, we have proposed a set of saddles which are points in the $A_2$ co-root lattice.  With this prescription for $\mathfrak{m}$, we calculate the partition function and observe that the divergence is indeed worse. Observing the trend of the divergence reflected on the the volume prefactor, we have also predicted a conjectural form for arbitrary higher spin cases. The dependence of the individual partition function on each Lens space scales as positive spin dependent power law. In contrast, in the purely bosonic theory this dependence was a negative power law, which made a concrete proof of convergence result for spins greater than 3, possible.

Apart from an immediate future problem as pointed above, as further future direction, we set aside the task of evaluating the quantum gravity partition function for the $\mathcal{N}=2$ supergravity theory, instead of the above purely Einstein gravity using the CS formulation. It would be interesting to study how the fermionic contributions from the supergravity theory differ from the present case. That eventually will be a valuable progress in classifying all possible excitations consistent with quantum gravity on de Sitter static patch.

\section*{Acknowledgements}

It is a pleasure to thank Nilay Kundu, Justin David and Rajesh Gupta for inputs and suggestions. RB acknowledges gratefully inputs from the participants of the workshop on ``Holography, Entanglement and Complexity" and hospitality provided by Ashoka University during the workshop. AR acknowledges various inputs from Arnab Kundu and Dharmesh Jain on various matters. He also is thankful for the hospitality extended by IISc, Bengaluru and BITS, Goa Campus during various stages of this work. And lastly, and as always, AR also expresses thanks to Al and Pat. 

RB acknowledges support by Fulbright foundation, DST(India) Inspire Award and OPERA Award from BITS, Pilani. 
AR is supported by Department of Atomic Energy (DAE), Government of India. 

\section{Appendix} \label{appendix} \subsection*{Conventions, Definitions, Notations and Identities}

Curved (World) indices = $\left\lbrace \mu,\nu,\sigma, ... \right\rbrace $.\qquad
Flat (Local Lorentz) indices = $ \left\lbrace I,J,K, ...\right\rbrace $. \nl
Frame fields = $\left\lbrace e^I_{\mu}\right\rbrace $, \quad Spin connections = $\left\lbrace \omega^{IJ}_\mu\right\rbrace $, \quad Connection 1-form = $\left\lbrace \omega^{IJ} \equiv \omega^{IJ}_\mu ~ dx^\mu \right\rbrace $. \nl
$\epsilon_{123}=\epsilon^{123}=+1$, \qquad $e^I\equiv e^I_\mu ~ dx^\mu$ , \qquad $\omega^I \equiv \frac{1}{2}\epsilon^{IJK}\omega_{JK}$ . \nl
$f_{IJK}= f_{[IJK]}$\footnote{Complete anti-symmetrization of the structure constants holds since the group in consideration is a compact Lie group. For non-compact gauge groups, this does not hold.}, \qquad $\epsilon_{IJK}\epsilon^{ILM}=+(\delta_J^L\delta_K^M-\delta_K^L\delta_J^M)$, \qquad a$_{[n_1 n_2]} \equiv\frac{1}{2!}(a_{n_1 n_2}-a_{n_2 n_1})$. \nl
Gauge Group $\equiv G$, \qquad Lie Algebra of $G \equiv \mathfrak{g}$, \qquad Cartan Sub-Algebra $\equiv \mathfrak{h}$.
\newpage
\bibliography{BibFile}
\end{document}